\documentclass[
 reprint,
superscriptaddress,
 amsmath,amssymb,
 aps,
]{revtex4-1}
\usepackage{graphicx}
\graphicspath{ {images/} }
\usepackage{dcolumn}
\usepackage{bm}
\usepackage{color,soul}
\usepackage[pdftex,linkcolor=blue,citecolor=blue,urlcolor=blue,colorlinks]{hyperref}
\usepackage[dvipsnames]{xcolor}
\usepackage{graphicx,epsfig}
\usepackage{color}
\usepackage{cancel}
\usepackage{dsfont}
\usepackage{inputenc}
\usepackage{amsmath}
\usepackage{amsfonts}
\usepackage{fancyhdr}
\usepackage{ulem}

\newcommand{\ket}[1]{\ensuremath{\left| #1 \right\rangle}}

\newcommand{\fref}[1]{Fig.~\ref{#1}}

\newcommand{\sref}[1]{Sec.~\ref{#1}}

\newcommand{\cref}[1]{chapter~\ref{#1}}
\newcommand{\Cref}[1]{Chapter~\ref{#1}}

\begin{document}

\preprint{APS/123-QED}

\title{Few-body bound topological and flat-band states in a Creutz Ladder}

\author{G. Pelegrí}
\affiliation{Department of Physics and SUPA, The University of Strathclyde, John Anderson Building, 107 Rottenrow East, Glasgow G4 0NG, UK}
\author{S. Flannigan}
\affiliation{Department of Physics and SUPA, The University of Strathclyde, John Anderson Building, 107 Rottenrow East, Glasgow G4 0NG, UK}
\author{A. J. Daley}
\affiliation{Department of Physics and SUPA, The University of Strathclyde, John Anderson Building, 107 Rottenrow East, Glasgow G4 0NG, UK}
\affiliation{Clarendon Laboratory, University of Oxford, Parks Road, Oxford OX1 3PU, United Kingdom}


\begin{abstract}
We investigate the properties of few interacting bosons in a Creutz ladder, which has become a standard model for topological systems, and which can be realised in experiments with cold atoms in optical lattices. At the single-particle level, this system may exhibit a completely flat energy landscape with non-trivial topological properties. In this scenario, we identify topological two-body edge states resulting from the bonding of single-particle edge and flat-band states. We also explore the formation of two- and three-body bound states in the strongly-interacting limit, and we show how these quasi-particles can be engineered to replicate the flat-band and topological features of the original single-particle model. Furthermore,  we show that in this geometry perfect Aharonov-Bohm caging of two-body bound states may occur for arbitrary interaction strengths, and we provide numerical evidence that the main features of this effect are preserved in an interacting many-body scenario resulting in many-body Aharonov-Bohm caging. 
\end{abstract}

\pacs{Valid PACS appear here}
\maketitle

\section{Introduction}
Over the last decades,  the study of topological materials has emerged as a central topic in condensed matter physics.  At the single-particle level, topological band theory provides a complete description of the properties of topological systems via the calculation of global order parameters \cite{TopInsReview,reviewTopIns2,TopInsSymReview}. However, while it is known that underlying topological band structures are responsible for the appearance of striking many-body phases such as fractional quantum Hall states \cite{QuantumHallFrac}
 or symmetry-protected topological phases \cite{SPTSenthil}, there are still many open questions regarding the interplay between topology and interactions. In an effort to build a bottom-up understanding of interacting topological models, a number of studies have focused on the properties of systems with a small number of particles \cite{DoublonHarperHofstadter,SSH2bosons1,SSH2bosons2,SSHHolesChain1,SSHHolesChain2,SSH2particlesRing,SSH2bosons3,
Haldane2bosons,Topology2photons1,Topology2photons2,TopologyFermionicDoublons1,
TopologyFermionicDoublons2,DoublonLattices,TwoBodyCreutz,Ke-2017,Lin-2020,Lyubarov-2019,stepanenko2020interaction,azcona2021doublons,stepanenko2021probing, pelegri2020interaction,kuno2020interaction}. Such few-body states can be prepared and probed in a number of experimental platforms such as ultracold atoms \cite{ReviewUltracoldTop} (which also allow for controllable interaction strengths),  photonic systems \cite{ReviewTopologicalPhotonics} or topolectrical circuits \cite{TwoBodyTopologyTopolectrical}.
\\
\\
Here we consider a small number of interacting bosonic particles in a Creutz ladder \cite{CreutzLadder}, which has been shown to be realisable with ultracold atoms \cite{Mugel2014,zhang2014,similar2,Song2018,kang2020,flannigan2020enhanced}. For a single particle this model possesses non-trivial topological properties which lead to the appearance of symmetry-protected edge states \cite{Zurita2020topology}. Making use of analytical calculations and perturbation theory, we identify several regimes in which these topological signatures are carried over bound states of two \cite{zheng2023two} and three particles.  Besides its general topological nature, the Creutz ladder is interesting because for appropriate values of the magnetic flux it exhibits a completely flat band structure characterised by Compact Localized States (CLS) arising from quantum interference \cite{ReviewFlatBands}.  Due to the suppression of the kinetic energy term, under the presence of even very weak interactions flat band structures give rise to a plethora of exotic many-body phases \cite{TopoFlatBands1,TopoFlatBands2,TopoFlatBands3,TopoFlatBands4,similar2,PhysRevB.106.014504} including pair superfluids \cite{flannigan2020enhanced} or many-body localized states \cite{kuno2020extended,orito2021interplay}.  

We begin here by investigating the limit in which the single-particle spectrum is composed of flat bands, identifying the existence of unusual two-body edge states formed by the bonding of a single-particle topological edge state and a CLS localized at the edge of the ladder. These states exist for all values of the two-body interaction strength and are present away from the strict flat-band limit. In the strongly interacting limit, it is possible to derive an effective model for two-body bound states consisting of the two particles occupying the same site. These quasi-particles, commonly referred to as doublons, are found to replicate the same physics as the original Creutz ladder with renormalized parameters, and contain topological and non-topological edge states in their spectrum. We generalize the procedure used to derive the doublon model to find an effective description for three-body bound states, which we refer to as trions. Similarly to the two-body case, we find that trions behave in an analogous way to the single particles, and have a clearer distinction between topological and non-topological edge states than doublons. 

For a Creutz ladder in the absence of interactions, a flat energy spectrum leads to Aharonov-Bohm caging \cite{ABcageoriginal,diamondchain1,diamondchain2,recentwork1,recentwork2, ABcagingPhotonics}, which consists on the dynamical localization of wavepackets on a small region of the system determined by the CLS.  However, the presence of interactions generally disrupts this effect and leads to leakage outside of the cage \cite{ABinter2,ABinter1,ABinter2part1,ABinter2part2}. Here we show that the Creutz ladder can support perfect Aharonov-Bohm caging for two-body bound states \cite{pelegri2020interaction} for arbitrary interaction strengths, and that the main signatures of this effect are retained in the many-body scenario \cite{PhysRevA.107.023305}. In the Creutz ladder it is possible to have perfect caging of two interacting particles for arbitrarily high interactions by preparing a specific initial state, offering potential applications in the building of atomtronic circuits \cite{amico2021roadmap}. Using Matrix Product State calculations, we verify that this effect is qualitatively retained for many-body states. We also show that arbitrary initial states composed of doublons localized in different parts of the lattice retain the main signatures of Aharonov-Bohm caging in the strongly interacting limit.
\\
\\
The rest of paper is organized as follows. In \sref{twobody_sec} we focus on the properties of two-body bound states, which we generalize to three-body bound states in \sref{3bodyBound}.  In \sref{ABcaging_sec}, we then address different types of many-body Aharonov-Bohm caging that may occur in the system, and finally in \sref{conclusion} we summarize the main conclusions of this work.

\section{Two-body interacting bound States}\label{twobody_sec}
\begin{figure}
\includegraphics[width=\linewidth]{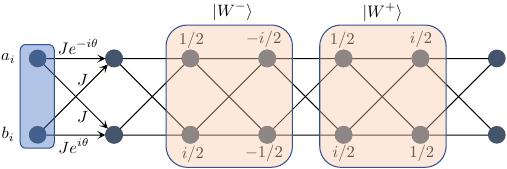}\\
\caption{The Creutz ladder geometry considered in this work, with the unit cell composed of an $a$ and a $b$ site highlighted in blue. The orange boxes depict the region of support for the Wannier functions associated with the lowest $\left(\ket{W^-}\right)$ and highest $\left(\ket{W^+}\right)$ flat bands that occur for $\theta=\pi/2$.}
\label{C_Lad}
\end{figure}

The Creutz ladder geometry that we consider is illustrated in \fref{C_Lad}. Each unit cell is formed by two sites, denoted as $a$ and $b$,  with horizontal and diagonal tunneling rate $J$, and a total flux per plaquette $2\theta$. The lattice is loaded with bosons with a two-body interaction strength $U$,  and therefore the many-body Hamiltonian of the system is given by 
\begin{equation}\label{TB_MOD}
\begin{split}
& \hat{H}  =  \frac{U}{2}\sum_{i=1}^{N_c} (\hat{a}_i^{\dagger} \hat{a}_i^{\dagger} \hat{a}_i \hat{a}_i + \hat{b}_i^{\dagger}\hat{b}_i^{\dagger} \hat{b}_i \hat{b}_i)\\
&+J\sum_{i=1}^{N_c-1} \left[\hat{a}_i^{\dagger}\hat{b}_{i+1}+\hat{b}_i^{\dagger}\hat{a}_{i+1}+e^{i\theta}\hat{a}_{i}^{\dagger}\hat{a}_{i+1}+e^{-i\theta}\hat{b}_{i}^{\dagger}\hat{b}_{i+1}+h.c.\right],
\end{split}
\end{equation}
where $N_c$ is the total number of unit cells, and $\hat{a}_i^{\dagger}$  ($\hat{b}_i^{\dagger}$) creates a particle on the $a$ ($b$) site of the $i$th unit cell.  For $\theta\neq 0$(mod $2\pi$), the single-particle band structure of this model is topologically non-trivial, with each of the two bands having a Zak's phase $\mathcal{Z}=\pi$ \cite{Zurita2020topology}. According to the bulk-boundary correspondence, in this topological phase a ladder with open boundary conditions has edge states protected by the inversion symmetry of the ladder. 
\subsection{Single-particle flat-band limit}
\begin{figure}[t!]
\includegraphics[width=\linewidth]{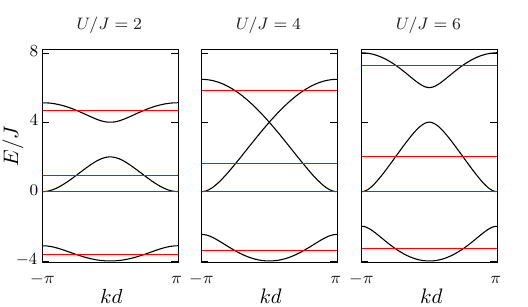}
\centering
\caption{Band structure of two-boson bound states in the Creutz Ladder for the case $\theta=\pi/2$ for two-body interaction strengths $U/J=2,4,6$. The black lines are the energy bands resulting from diagonalization of the Hamiltonian \eqref{Hk_doublons_1}, while the red lines are the energies corresponding to the eigenstates of the Hamiltonian \eqref{Hk_doublons_2}.}
\label{doublonbands}
\end{figure}

\begin{figure}
\includegraphics[width=\linewidth]{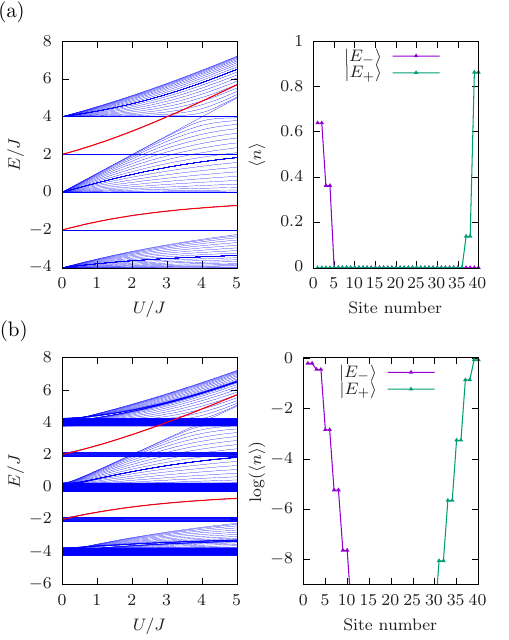}
\caption{Left plots: Two-boson energy spectrum of an open Creutz ladder formed by 20 unit cells as a function of the interaction strength $U$ for a hopping phase (a) $\theta=\pi/2$ and (b) $\theta=0.48\pi$. Right plots: density distribution of the lower and higher energy edge states for $U=2J$ and a hopping phase (a) $\theta=\pi/2$ and (b) $\theta=0.48\pi$. The sites are labelled as $a_1,a_2,a_3=1,3,5,...;b_1,b_2,b_3=2,4,6,...$, i.e., odd numbers are assigned to $a$ type sites and even numbers to $b$ type sites.}
\label{Spectrum_EdgeStates}
\end{figure}

The single-particle spectrum of the Creutz ladder becomes particularly interesting for $\theta=\pi/2$. In this case,  the system has two completely flat bands of energies $E=\pm 2J$ and the Wannier functions associated to each flat band are completely localized in only two unit cells, as shown in \fref{C_Lad}. The creation operators for the higher/lower band states, $\hat{W}_i^{\pm\dagger} $, can be expressed in terms of the operators for the $a$ and $b$ sites as
\begin{align}
W_i^{+\dagger}&=\frac{1}{2}\hat{a}_{i}^{\dagger}+\frac{i}{2}\hat{b}_{i}^{\dagger}+\frac{i}{2}\hat{a}_{i+1}^{\dagger}+\frac{1}{2}\hat{b}_{i+1}^{\dagger}\\
W_i^{-\dagger}&=\frac{1}{2}\hat{a}_{i}^{\dagger}+\frac{i}{2}\hat{b}_{i}^{\dagger}-\frac{i}{2}\hat{a}_{i+1}^{\dagger}-\frac{1}{2}\hat{b}_{i+1}^{\dagger}.
\end{align}
Additionally, for open boundary conditions the leftmost and righmost unit cells host each one edge state of $E=0$. The creation operators $\hat{L}^{\dagger}$ and $\hat{R}^{\dagger}$ of these states read
\begin{align}
\hat{L}^{\dagger}&=\frac{1}{\sqrt{2}}\hat{a}_1^{\dagger}-\frac{i}{\sqrt{2}}\hat{b}_1^{\dagger},\\
\hat{R}^{\dagger}&=\frac{i}{\sqrt{2}}\hat{a}_N^{\dagger}-\frac{1}{\sqrt{2}}\hat{b}_N^{\dagger}.
\end{align}
The basis formed by the Wannier functions and the edge states is particularly convenient for analysing the bound states. In this basis, the Hamiltonian of a Creutz ladder with open boundary conditions is given by
\begin{align} 
& \hat{H} = \hat{H}_{\text{bulk}}+\hat{H}_{\text{edges}};\label{Int}\\
& \hat{H}_{\text{bulk}}=-2J\sum_{i=1}^{N_c-1} \hat{W}_i^{-\dagger} \hat{W}_i^+ + 2J \sum_{i=1}^{N_c-1}  \hat{W}_i^{+\dagger} \hat{W}_i^+\nonumber\\
& +  \frac{U}{16} \sum_{i=2}^{N_c-2} \left[ \left( \hat{W}_i^{\dagger}\hat{W}_i^{\dagger}\hat{W}_i\hat{W}_i + \tilde{W}_i^{\dagger}\tilde{W}_i^{\dagger} \tilde{W}_i \tilde{W}_i \right) \right. \nonumber\\
& \left. + \left(4\hat{W}_i^{\dagger}\hat{W}_i \tilde{W}_{i+1}^{\dagger}\tilde{W}_{i+1} - \hat{W}_i^{\dagger}\hat{W}_i^{\dagger} \tilde{W}_{i+1} \tilde{W}_{i+1} + h.c\right) \right],\label{Int_Bulk}\\
& \hat{H}_{\text{edges}}=\frac{U}{4}\left(\hat{L}^{\dagger}\hat{L}^{\dagger}\hat{L}\hat{L}+\hat{R}^{\dagger}\hat{R}^{\dagger}\hat{R}\hat{R}\right)\nonumber\\
& +\frac{U}{16}\left(\hat{W}_1^{\dagger}\hat{W}_1^{\dagger}\hat{W}_1\hat{W}_1+\tilde{W}_{N_c-1}^{\dagger}\tilde{W}_{N_c-1}^{\dagger}\tilde{W}_{N_c-1}\tilde{W}_{N_c-1}\right)\nonumber\\
& + \frac{U}{2}\left(\hat{W}_1^{\dagger}\hat{L}^{\dagger}\hat{W}_1\hat{L}+\tilde{W}_{N_c-1}^{\dagger}\hat{R}^{\dagger}\tilde{W}_{N_c-1}\hat{R}\right)\nonumber\\
& +\frac{U}{8}\left(\hat{L}^{\dagger}\hat{L}^{\dagger}\hat{W}_1\hat{W}_1+\hat{R}^{\dagger}\hat{R}^{\dagger}\tilde{W}_{N_c-1}\tilde{W}_{N_c-1}+h.c.\right),\label{Int_Edges}
\end{align}
where $\hat{W}_i = \hat{W}_i^{+} +\hat{W}_i^{-} $ and $\tilde{W}_i = \hat{W}_i^{+} - \hat{W}_i^{-} $. 
Due to the onsite interactions, the bulk Hamiltonian contains terms that couple single-particle Wannier states localized in neighbouring cells in different ways.  In particular, the final two terms describe the dynamics of two particles simultaneously hopping from one plaquette to an adjacent one, breaking the single-particle localization. Additionally, the term proportional to $\hat{W}_i^{\dagger}\hat{W}_i \tilde{W}_{i+1}^{\dagger}\tilde{W}_{i+1}$ yields an interaction energy between particles occupying neighbouring cells.  Taking these coupling terms into account, we can analyze the spectrum of two-boson bound states in an infinite ladder by considering two types of quasiparticles that do not mix with each other \cite{flannigan2020enhanced}.  The first type of composite particles is formed by states in which the two bosons occupy single-particle Wannier states localized in the same plaquette, i.e., two-boson states created by the combinations of Wannier-state operators $\{\frac{1}{\sqrt{2}}\hat{W}_i^{-\dagger}\hat{W}_i^{-\dagger}, \frac{1}{\sqrt{2}}\hat{W}_i^{+\dagger}\hat{W}_i^{+\dagger}, \quad \hat{W}_i^{+\dagger}\hat{W}_i^{-\dagger}\}$.  The terms in $\hat{H}_{\text{bulk}}$ that give rise to pair tunneling couple these three states,  and the momentum-space Hamiltonian of this subset of states reads
\begin{widetext}
\begin{equation}
H_k^1=\begin{pmatrix}
-4J+U/4(1-\cos kd)& U/4(1-\cos kd) & iU\sqrt{2}/4 \sin kd \\
U/4(1-\cos kd) & 4J+U/4(1-\cos kd) & iU\sqrt{2}/4 \sin kd \\
-iU\sqrt{2}/4 \sin kd & -iU\sqrt{2}/4 \sin kd & U/2(1-\cos kd)
\end{pmatrix},
\label{Hk_doublons_1}
\end{equation}
\end{widetext}
where $d$ is the separation between unit cells and we have used the basis ordering mentioned above. The three bands formed by these bound states are depicted with black lines in \fref{doublonbands}. For $U=4J$, the second and third bands form a Dirac cone around $kd=0$, where a gap closing occurs. For any other value of $U$ the three bands remain gapped. The second type of two-body bound states consists of each of the two particles occupying Wannier states localized on nearest neighbour plaquettes, and are therefore created by combinations of the operators $\{\hat{W}_i^{-\dagger}\hat{W}_{i+1}^{-\dagger}, \quad \hat{W}_i^{+\dagger}\hat{W}_{i+1}^{+\dagger}, \quad \hat{W}_i^{+\dagger}\hat{W}_{i+1}^{-\dagger}, \quad \hat{W}_i^{-\dagger}\hat{W}_{i+1}^{+\dagger}\}$. Since $\hat{H}_{\text{bulk}}$ does not contain any term describing single-particle hopping, these bound states form a closed system at each pair of plaquettes $i,i+1$, giving rise to a $k$-independent momentum space Hamiltonian
\begin{equation}
H_k^2=\begin{pmatrix}
-4J+U/4&-U/4&U/4&-U/4\\
-U/4&4J+U/4&-U/4&U/4\\
U/4&-U/4&U/4&-U/4\\
-U/4&U/4&-U/4&U/4
\end{pmatrix}.
\label{Hk_doublons_2}
\end{equation}
The four quasiparticle flat bands corresponding to this subspace are depicted as red lines in \fref{doublonbands}. 
\\
\\
In a finite ladder with open boundary conditions, there are additional bound states that result from the coupling between the Wannier states created by the operators $\hat{W}_{1}^{\pm\dagger}$ and $\hat{W}_{N_c-1}^{\pm\dagger}$, and the edge states created by $\hat{L}^{\dagger}$ and $\hat{R}^{\dagger}$, as described by the edge Hamiltonian \eqref{Int_Edges}. Examining the different terms of $\hat{H}_{\text{edges}}$, we notice that on the left edge of the ladder the two-boson states $\hat{L}^{\dagger}\hat{W}_1^{-\dagger}\ket{0}$ and $\hat{L}^{\dagger}\hat{W}_1^{+\dagger}\ket{0}$ are coupled between themselves but not to any other bound pairs, and the same occurs for  $\hat{R}^{\dagger}\hat{W}_{N_c-1}^{-\dagger}\ket{0}$ and $\hat{R}^{\dagger}W_{N_c-1}^{+\dagger}\ket{0}$ at the right edge.  Explicitly, the only non-zero matrix elements between states of this kind localized at the left edge are
\begin{align*}
&\langle0|\hat{L}\hat{W}_1^{-}\hat{\hat{H}}\hat{L}^{\dagger}\hat{W}_1^{-\dagger}|0\rangle=-2J+\frac{U}{2}\\
&\langle0|\hat{L}\hat{W}_1^{+}\hat{\hat{H}}\hat{L}^{\dagger}\hat{W}_1^{-\dagger}|0\rangle=\langle0|\hat{L}\hat{W}_1^{-}\hat{\hat{H}}\hat{L}^{\dagger}\hat{W}_1^{+\dagger}|0\rangle=\frac{U}{2}\\
&\langle0|\hat{L}\hat{W}_1^{+}\hat{\hat{H}}\hat{L}^{\dagger}\hat{W}_1^{+\dagger}|0\rangle=2J+\frac{U}{2}
\end{align*}
Similarly, the non-zero matrix elements between states localized at the right edge read
\begin{align*}
&\langle0|\hat{R}\hat{W}_{N_c-1}^{-}\hat{H}\hat{R}^{\dagger}\hat{W}_{N_c-1}^{-\dagger}|0\rangle=-2J+\frac{U}{2}\\
&\langle0|\hat{R}\hat{W}_{N_c-1}^{+}\hat{H}\hat{R}^{\dagger}\hat{W}_{N_c-1}^{-\dagger}|0\rangle=\langle0|\hat{L}\hat{W}_1^{-}\hat{H}\hat{L}^{\dagger}\hat{W}_1^{+\dagger}|0\rangle=-\frac{U}{2}\\
&\langle0|\hat{R}\hat{W}_{N_c-1}^{+}\hat{H}\hat{R}^{\dagger}\hat{W}_{N_c-1}^{+\dagger}|0\rangle=2J+\frac{U}{2}
\end{align*}
Writing these two systems of equations in matrix form and diagonalizing them, we find that the eigenstates at both edges have energies 
\begin{equation}
E_{\pm}=\frac{1}{2}\left( U\pm \sqrt{U^2+16J^2}\right).
\label{energies_edge}
\end{equation}
These states are strictly localized on the unit cells 1 and 2 (or $N_c-1$ and $N_c$), as they are composed of two-body states in which one particle occupies an edge state and the other an adjacent flat-band state.  In the right plot of \fref{Spectrum_EdgeStates} (a) we show the density distribution of the lower and higher energy edge states of an open Creutz ladder formed by 20 unit cells and with hopping phase $\theta=\pi/2$,  for an interaction strength $U/J=2$. In the left plot of \fref{Spectrum_EdgeStates} (a) we show the two-boson energy spectrum of the same ladder as a function of the interaction strength $U$. The lines coloured in red correspond to the two-boson edge states formed by the bonding of a single-particle edge state with a flat-band state. As predicted by the analytical values \eqref{energies_edge} of the edge-state energies, the higher-energy two-boson edge states become degenerate with the bulk for $U/J\geq 3$, but they remain localized at the edges of the ladder.  

Although the exact expressions for two-boson edge states formed by the bonding of a single-particle edge state and a flat-band state can only be found for $\theta=\pi/2$, these states are also present away from the single-particle flat band limit. In the upper plot of \fref{Spectrum_EdgeStates} (b) we show the two-boson spectrum of an open Creutz ladder formed by 20 unit cells with hopping phase  $\theta=0.48\pi$ as a function of $U$. While the bulk spectrum differs from the $\theta=\pi/2$ case depicted in \fref{Spectrum_EdgeStates} (a), the energies of the two-boson edge states (coloured in red) still have the dependence on the interaction strength given by eq. \eqref{energies_edge}.  In the lower plot of  \fref{Spectrum_EdgeStates} (b) we show the density distribution of the lower and higher energy edge states for the same ladder as in the upper plot and $U/J=2$.  Although the states are still largely localized on the two leftmost (righmost) unit cells, the density profile decays exponentially into the bulk. For larger deviations from the flat-band limit the edge localization becomes weaker.  

\subsection{Strongly-interacting limit}
Although the exact analytical study of the properties of two-boson bound states is only possible for the single-particle flat-band case, for sufficiently strong interactions $U\gg J$ it is possible to derive an effective model for high energy bound states, commonly referred to as doublons. In this limit, the formation of bound states occurs for any value of the ladder parameters, and is caused by the fact that, due to the finite bandwith of the single-particle spectrum, states with doubly occupied sites have a much higher energy than those with single site occupation, and therefore the two types of states do not mix.  Taking profit of the energy separation between the doublon bands and the rest of the spectrum, it is possible to perform a Schrieffer-Wolff transformation to find an effective Hamiltonian restricted to the doublon subspace, as we discuss in detail in appendix \ref{AppendixSW}. As we show in the appendix and in \sref{3bodyBound},  this procedure can be generalized to obtain effective models for bound states formed by more than two particles.  Defining the doublon annihilation operators $\hat{A}_i\equiv \frac{1}{\sqrt{2}}\hat{a}_i\hat{a}_i,\hat{B}_i\equiv \frac{1}{\sqrt{2}}\hat{b}_i\hat{b}_i$, we can write the effective Doublon Model (DM) Hamiltonian as
\begin{align}
&\hat{H}_{\text{eff}}^{N=2}=\frac{2J^2}{U}\sum_{i=1}^{N_c-1} \left(\hat{A}_i^{\dagger}\hat{B}_{i+1}+\hat{B}_i^{\dagger}\hat{A}_{i+1}\right.\nonumber\\
&\left.+e^{2i\theta}\hat{A}_{i}^{\dagger}\hat{A}_{i+1}+e^{-2i\theta}\hat{B}_{i}^{\dagger}\hat{B}_{i+1}+h.c.\right)\nonumber\\
&+\left(U+\frac{8J^2}{U}\right)\sum_{i=1}^{N_c}\left( \hat{A}_i^{\dagger}\hat{A}_i+\hat{B}_i^{\dagger}\hat{B}_i\right)\nonumber\\
&-\frac{4J^2}{U}\left(\hat{A}_1^{\dagger}\hat{A}_1+\hat{B}_1^{\dagger}\hat{B}_1+\hat{A}_{N_c}^{\dagger}\hat{A}_{N_c}+\hat{B}_{N_c}^{\dagger}\hat{B}_{N_c}\right).
\label{Ham_doublons_Creutz}
\end{align}

\begin{figure}[t!]
\includegraphics[width=\linewidth]{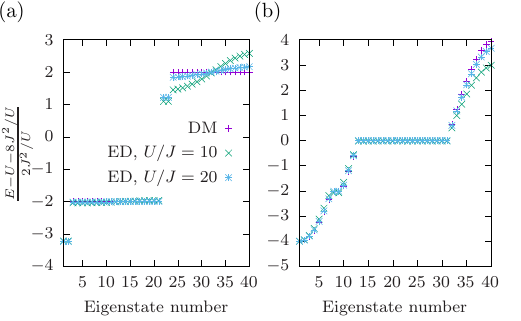}
\caption{Doublon spectra obtained through ED and the effective DM Hamiltonian \eqref{Ham_doublons_Creutz} for different values of $U$ in a Creutz ladder with open boundary conditions formed by 20 unit cells. The hopping phase is $\theta=\pi/4,3\pi/4$ in plot (a), and $\theta=\pi/2$ in plot (b).}
\label{Energies_Eff_ED_Doublons}
\end{figure}

\begin{figure}[t!]
\includegraphics[width=\linewidth]{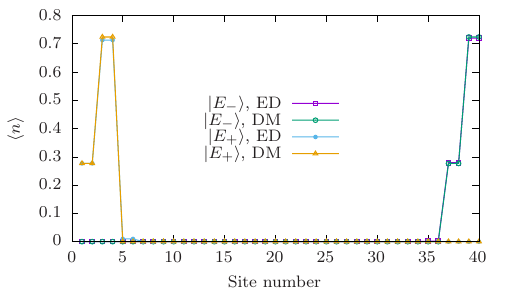}
\caption{Density profiles of the lower energy and higher energy doublon edge state obtained with ED and the effective DM Hamiltonian \eqref{Ham_doublons_Creutz} for $U=20J$ in a Creutz ladder with open boundary conditions formed by 20 unit cells and with hopping phase $\theta=\pi/4,3\pi/4$.}
\label{EdgestatesDoublons}
\end{figure}
The first two terms describe an effective Creutz ladder with renormalized tunneling rate $J\rightarrow \frac{2J^2}{U}$ and with a renormalized complex hopping phase $\theta\rightarrow 2\theta$. The third term is an overall energy shift, coming from second order processes in which a doublon splits and then recombines in the same site. Finally, the last term, which only appears for open boundary conditions, is an effective energy shift on the four edge sites stemming from the fact that they have different coordination numbers than the bulk sites. 

For $U\gg J$, we expect the effective DM \eqref{Ham_doublons_Creutz} to reproduce well the high-energy sector of the spectrum obtained by full Exact Diagonalization (ED) of a Creutz ladder with 2 bosons. As we show in \fref{Energies_Eff_ED_Doublons}, where we plot the doublon energies computed with full ED and the effective model in an open ladder formed by 20 unit cells for $\theta=\pi/4,3\pi/4$ (a),  $\theta=\pi/2$ (b) and different values of $U$,  the convergence between the two methods is better for higher interaction strengths. By analogy with the solution of a single-particle in the Creutz ladder,  for $\theta=\frac{\pi}{4}, \frac{3\pi}{4}$ we expect the bulk doublon spectrum to be formed by two flat bands, as can be observed in the upper plot of \fref{Energies_Eff_ED_Doublons}.  In a ladder with open boundary conditions, since the edge sites have an energy shift of $-4J^2/U$ relative to the the bulk sites, we expect to observe isolated edge states in the doublon spectrum originated by this energy shift. However, since this shift is only two times larger than the effective hopping ($2J^2/U$), these states will also have some admixture from the closest cells to the edges. Besides these non-topological edge states, by analogy with the single-particle solution we also expect to observe topological doublon edge states with an energy laying in the gap between the bulk bands. Since the edge sites are shifted in energy, these edge states will tend to be displaced to the unit cells number $2$ and $N_c-1$, but will still partially hybridize with the non-topological edge states. 

As a result of this interplay between the two different types of edge states, in \fref{Energies_Eff_ED_Doublons} (a) we observe two different pairs of edge states. The first type of edge states, which we denote $\ket{E-}$ and lay at a lower energy than the first doublon bulk band, have a higher admixture of non-topological edge states, and their density is therefore more concentrated on the unit cell 1 (or $N_c$), as shown in \fref{EdgestatesDoublons}. The second type of edge states, which we denote $\ket{E+}$ and lay in the  gap between the two doublon bulk bands, have a higher admixture of the topological edge states, and are therefore more localized on the unit cell 2 (or $N_c$), as shown in \fref{EdgestatesDoublons}.  As shown in \fref{Energies_Eff_ED_Doublons} (b), for $\theta=\pi/2$ the doublon bulk spectrum is gapless and the energy of the edge states falls inside the bulk band. This case corresponds precisely to the single-particle flat-band limit studied in the previous section. Therefore, the two edge states merged with the lower bulk band that can be observed in the plot can be interpreted as the bonding of a single-particle edge state and a flat-band state, as discussed previously.  

\section{Three-body bound states}
\label{3bodyBound}
We now consider a Creutz ladder filled with 3 interacting bosons.  In the strongly interacting limit $U\gg J$ we can apply the Schrieffer-Wolff transformation described in appendix \ref{AppendixSW} to derive an effective model for interacting three-body bound states. Defining the three-body annihilation operators $A_i\equiv \frac{1}{\sqrt{6}}a_ia_ia_i,B_i\equiv \frac{1}{\sqrt{6}}b_ib_ib_i$, we can write the effective Trion Model (TM) Hamiltonian as 
\begin{align}
&\hat{H}_{\text{eff}}^{N=3}=\frac{6J^3}{4U^2}\sum_{i=1}^{N_c-1}\left( \hat{A}_i^{\dagger}\hat{B}_{i+1}+\hat{B}_i^{\dagger}\hat{A}_{i+1}\right.\nonumber\\
&\left.+e^{3i\theta}\hat{A}_{i}^{\dagger}\hat{A}_{i+1}+e^{-3i\theta}\hat{B}_{i}^{\dagger}\hat{B}_{i+1}+h.c.\right)\nonumber\\
&+\left(3U+\frac{12J^2}{2U}\right)\sum_{i=1}^{N_c} \hat{A}_i^{\dagger}\hat{A}_i+\hat{B}_i^{\dagger}\hat{B}_i\nonumber\\
&-\left(\frac{12J^2}{4U}\right)\left(\hat{A}_1^{\dagger}\hat{A}_1+\hat{B}_1^{\dagger}\hat{B}_1+\hat{A}_{N_c}^{\dagger}\hat{A}_{N_c}+\hat{B}_{N_c}^{\dagger}\hat{B}_{N_c}\right)
\label{Ham_trions_Creutz}
\end{align}
The interpretation of the effective TM Hamiltonian \eqref{Ham_trions_Creutz} is similar to the one of the effective doublon model given by eq. \eqref{Ham_doublons_Creutz}.  The first two terms describe an effective non-interacting Creutz ladder with renormalized tunneling rate $J\rightarrow \frac{6J^3}{4U^2}$ and with a renormalized complex hopping phase $\theta\rightarrow 3\theta$. The third term is an overall energy shift due to the interaction energy between three bosons, $3U$, and second order processes in which a trion is split and recombined on the same site. Finally, the last term is a relative energy shift on the four edge sites due to their reduced coordination numbers with respect to the bulk sites, and only appears for open boundaries.  

\begin{figure}[t!]
\includegraphics[width=\linewidth]{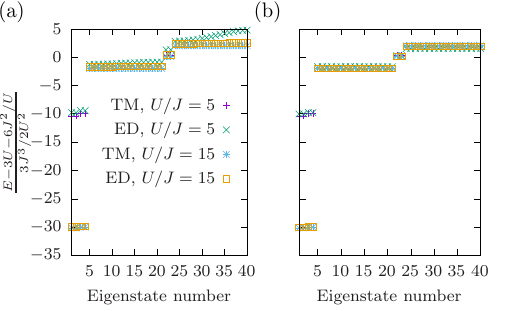}
\caption{Trion spectra obtained through ED and the effective TM Hamiltonian \eqref{Ham_trions_Creutz} for different values of $U$ in a Creutz ladder with open boundary conditions formed by 20 unit cells. The hopping phase is $\theta=\pi/3,5\pi/6$ in plot (a), and $\theta=\pi/2$ in plot (b).}
\label{Energies_Eff_ED_Trions}
\end{figure}

As we discussed in the case of doublons, this relative energy shift gives rise to non-topological edge states and causes a displacement by one unit cell of the topological ones. By analogy with the solution of a single-particle in the Creutz ladder, we expect to find trion flat bands for $\theta=\frac{\pi}{6},\frac{5\pi}{6}$ and $\frac{\pi}{2}$.  We note that the latter value of the hopping phase coincides with the one for which the single-particle spectrum is also formed by flat bands, so we expect the ED results for $\theta=\frac{\pi}{2}$ to differ slightly from the ones corresponding to $\theta=\frac{\pi}{6}, \frac{5\pi}{6}$, even though the effective TM Hamiltonian \eqref{Ham_trions_Creutz} is equivalent for all three values of $\theta$.  This is confirmed in \fref{Energies_Eff_ED_Trions}, where we plot a comparison between the trion energies computed with ED and the effective TM on an open Creutz ladder formed by 20 unit cells for different values of $U$,  and hopping phase $\theta=\frac{\pi}{6},\frac{5\pi}{6}$ (a) and $\theta=\frac{\pi}{2}$(b).  Due to the underlying single-particle flat-band spectrum, for $\theta=\frac{\pi}{2}$ the three-body bound states form completely flat bands for lower values of $U$ than in the case $\theta=\frac{\pi}{6},\frac{5\pi}{6}$, for which the strongly-interacting condition $U\gg J$ must be fulfilled in order for the flat bands predicted by the effective model to emerge.  

\begin{figure}[t!]
\includegraphics[width=\linewidth]{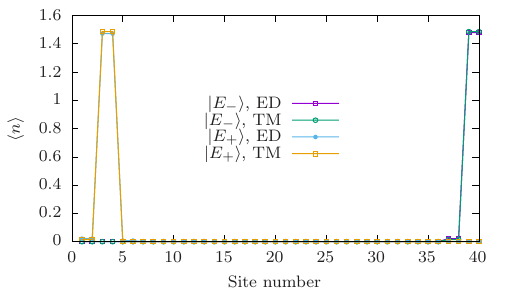}
\caption{Density profiles of the lower energy and higher energy trion edge states obtained with ED and the effective TM Hamiltonian \eqref{Ham_trions_Creutz} for $U=10J$ in a Creutz ladder with open boundary conditions formed by 20 unit cells and with hopping phase $\theta=\pi/2$.}
\label{EdgestatesTrions}
\end{figure}

In \fref{Energies_Eff_ED_Trions} we also observe that the gap between the lower energy edge states and the lower flat and increases with $U$. This is due to the fact that the energy shift of the edge sites is $\sim J^2/U$, while the effective trion hoppings are $\sim J^3/U^2$.  Due to the large energy difference between the edge site shift and the effective coupling, the admixture effect between topological and non-topological edge states that occurs in the doublon case is highly suppressed for trions. The consequences of this can be observed in \fref{EdgestatesTrions}, where we plot, for an open Creutz ladder formed by 20 unit cells with hopping phase $\theta=\pi/2$ and $U=10J$, the density profiles of the lower-energy and higher-energy trion edge states, computed with ED and the effective TM. The lower-energy states $\ket{E-}$, which are non-topological and due to the onsite shift on the edge sites, are highly localized on the edge sites of the ladder, while the higher-energy states $\ket{E+}$, which are of topological origin and lay in the gap between the two flat bands, are highly localized on the unit cell number 2 (or $N_c-1$).

\section{Aharonov-Bohm caging}
\label{ABcaging_sec}
\subsection{Two-body Aharonov-Bohm caging}
Aharonov-Bohm caging is a phenomenon consisting on the confinement of wave packets to a small region of the lattice \citep{ABcageoriginal}. At the single-particle level, this effect can be explained as a consequence of the suppression of the kinetic energy in flat-bands systems. In previous works it has been shown that interactions tend to hinder Aharonov-Bohm caging and cause the delocalization of particles in flat-band systems \cite{ABinter2,ABinter1,ABinter2part1,ABinter2part2}. However, as we explain below the Creutz ladder can host different types of two-body Aharonov-Bohm cages.  On the one hand, for $\theta=\pi/2$ it is possible to identify a particular set of initial states which exhibit perfect caging for arbitrary interactions strengths. On the other hand, as already pointed out in \cite{Zurita2020topology}, for $\theta=\pi/4,3\pi/4$ and in the strongly interacting limit the dynamics of doublons are governed by a Hamiltonian analogous to that of single particles, and thus exhibit Aharonov-Bohm caging. Both types of Aharonov-Bohm cages could find applications in the design of atomtronic circuits \cite{amico2021roadmap}. 
\subsubsection{$\theta=\pi/2$}
For $\theta=\pi/2$, the existence of the two-boson bound flat-band states in which each of the two particles occupies neighbouring compact localized Wannier states (i.e., the eigenstates of the Hamiltonian \eqref{Hk_doublons_2}) can be probed through the Aharonov-Bohm caging effect by preparing an experimentally accessible initial state and observing its dynamics. In particular, the doublon superposition state $\ket{\Psi_i}=\frac{1}{2}\left(\hat{a}_i^{\dagger}\hat{a}_i^{\dagger}+\hat{b}_i^{\dagger}\hat{b}_i^{\dagger}\right)\ket{0}$ can be expressed in terms of the Wannier basis as
\begin{align}
\ket{\Psi_i}&=\frac{i}{2}\left(\hat{W}_{i-1}^{-\dagger}\hat{W}_{i}^{-\dagger}+\hat{W}_{i-1}^{-\dagger}\hat{W}_{i}^{+\dagger}\right.\nonumber\\
&\left.-\hat{W}_{i-1}^{+\dagger}\hat{W}_{i}^{-\dagger}-\hat{W}_{i-1}^{+\dagger}\hat{W}_{i}^{+\dagger}\right)\ket{0}=\frac{1}{2i}\tilde{W}_{i-1}^{\dagger}\hat{W}_{i}^{\dagger}\ket{0}.
\label{2Bcat}
\end{align}
Therefore, if one initializes two bosons in $\ket{\Psi_i}$ the time evolution will consist of oscillations between the different bound states forming a closed system in the plaquettes $i-1$ and $i$, and the atomic density will remain confined to the sites forming these two plaquettes.  As illustrated in \fref{ABcaging} (a), where we plot the time evolution of the atomic density in the 10th cell of a Creutz ladder with $\theta=\pi/2$ formed by a total of 20 cells after preparing two bosons in the state $\ket{\Psi_{10}}$, this caging effect occurs for any value of the interaction strength $U$. 
\subsubsection{$\theta=\pi/4,3\pi/4$, strongly interacting limit}
\begin{figure}
\includegraphics[width=\linewidth]{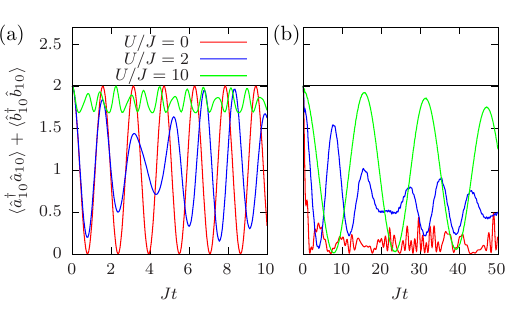}
\caption{Time evolution of the population of the 10th unit cell on an open Creutz ladder formed by 20 unit cells. The two bosons are initialized on the state $\ket{\Psi_{10}}$ as defined on the main text. The hopping phases are $\theta=\pi/2$ (a) and $\theta=\pi/4,3\pi/4$ (b).}
\label{ABcaging}
\end{figure}

As we discussed previously, in the strongly interacting limit the two highest energy doublon bands become flat when the hopping phase is $\theta=\pi/4,3\pi/4$. Therefore, in this situation we expect to find two-boson Aharonov-Bohm caging for any initial state, in the same way as it occurs for non-interacting states in the single-particle flat-band limit.  Since the flat-band maximally localized states extend over two unit cells, an initial state prepared in unit cell $i$ is confined to oscillate between cells $i-1$, $i$ and $i+1$. In \fref{ABcaging} (b) we plot the time evolution of the atomic density in the 10th cell of a Creutz ladder with $\theta=\pi/4,3\pi/4$ formed by a total of 20 cells after preparing two bosons in the state $\ket{\Psi_{10}}$. We observe that as the interaction strength $U$ is increased a higher fraction of the population remains confined in the cage. We also note that the frequency of the oscillations decreases with $U$ due to the fact that the gap between the two flat bands, given by $16J^2/U$ in the effective doublon model, becomes smaller. 

\subsection{Many-body Aharonov-Bohm caging}

\begin{figure}
\includegraphics[width=\linewidth]{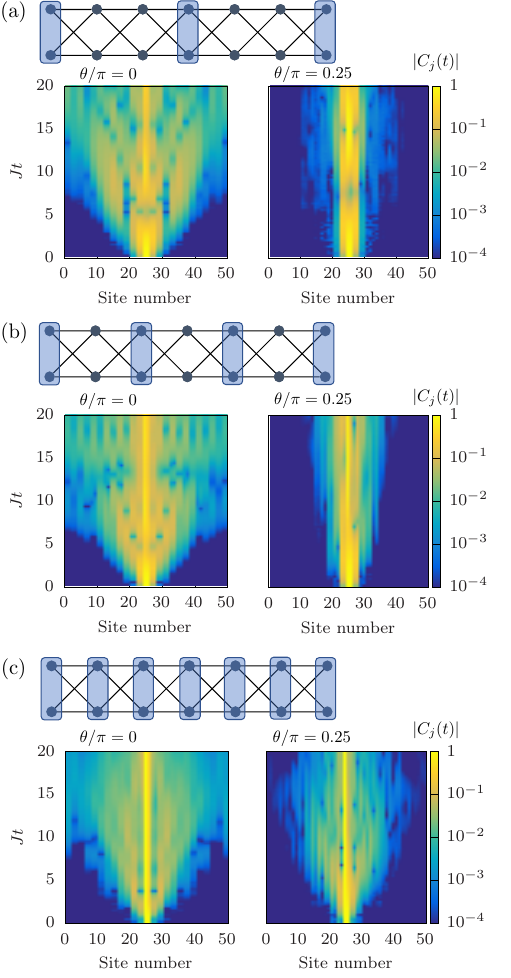}
\caption{Space-time evolution of two-body pair correlations (Eq.~\ref{MB_Corr_Eq}) in a Creutz ladder loaded with different densities of the two-boson states $\ket{\Psi_i}$ defined by eq. \eqref{2Bcat} for hopping phases $\theta=0$ and $\theta=\pi/4$. The initial state is constructed by loading the doublon superposition states \eqref{2Bcat} every three unit cells (a), every other unit cell (b) or in all unit cells (c). In all cases, we have considered an attractive interaction strength $U=-20J$ and $M = 15$ unit cells.}
\label{ABcaging_MB}
\end{figure}

Moving beyond the two-particle limit, another interesting question to examine is to which extent Aharonov-Bohm caging is preserved in an interacting many-body scenario.  In order to investigate this, in this section we use Matrix Product State methods to compute the dynamics of different initial states in a Creutz ladder formed by $M=25$ unit cells for hopping phases $\theta=0$ and $\theta=\pi/4$.  In order to maximize the contributions from doublons and minimize the corrections arising from trions, we work in the regime of strong attractive interactions $U<0, |U|\gg J$. In this limit, strong three-body losses occurring in experimental realizations of ultra-cold atoms in optical lattices~\cite{PhysRevResearch.2.043050,dincao2018JPB,braaten2006PRep} lead to an effective quantum Zeno type suppression of many-body states with more than two particles per site~\cite{Syassen1329,Yan:2013aa,Zhu2013,PhysRevA.82.022120}, and we can therefore introduce an upper bound of two in the single-site occupation of our model ~\cite{Schmidt2009,PhysRevB.84.092503,PhysRevA.81.061604,PhysRevLett.106.185302,PhysRevLett.103.240401,Daley2009,PhysRevLett.104.096803,PhysRevB.82.064509}. This effectively puts an upper bound on the local dimension of the numerical algorithm and gives us a potential experimental way of probing the effect of the unique features of doublons in this model on the many-body dynamics.

We have considered initial many-body states $\ket{\psi}$ constructed by loading the doublon superposition states given by Eq.~\eqref{2Bcat} in alternating unit cells with different densities, and computed the expectation value of the pairing correlation operator
\begin{equation} \label{MB_Corr_Eq}
C_j(t) = \langle \psi(t)| \hat{a}_{M/2}^{\dagger}\hat{a}_{M/2}^{\dagger}\hat{a}_{j}\hat{a}_{j} |\psi(t) \rangle,
\end{equation}
as a function of the rung position $j$ and time $t$.  The results are shown in Fig. \ref{ABcaging_MB}, where in all cases the interaction strength is $U=-20J$ so that we are in a regime where the lowest energy two-body band is effectively flat in the case $\theta=\pi/4$. Note that in all cases conventional single particle correlations, $\langle \hat{a}_{M/2}^{\dagger}\hat{a}_{j}\rangle$, remain exponentially suppressed for the timescales considered. In Fig. \ref{ABcaging_MB} (a) the initial state is constructed by loading two bosons into the doublon supersposition state $\ket{\Psi_i}$ every three unit cells.  In a scenario with perfect Aharonov-Bohm caging each doublon supersposition state would be confined to a cage formed by the cell in which it is prepared and the two adjacent ones, and therefore the correlations would not spread beyond these three unit cells. In practice, for a hopping phase $\theta=\pi/4$, for which we have a flat band in the two-body band structure, the two-body pair correlations spread further in the lattice due to the finite population leaking outside of the cage, but this spreading remains largely suppressed compared to the case with hopping phase $\theta=0$, for which there is significant single doublon dispersion.  

In Fig.~\ref{ABcaging_MB} (b) we consider an initial state in which the doublon supersposition state is prepared in every other unit cell. Even though in this case the ideal Aharonov-Bohm cages are overlapping, the correlation spreading for $\theta=\pi/4$ remains limited and clearly inferior to that observed for $\theta=0$. Finally, in Fig. \ref{ABcaging_MB} (c) we show the results corresponding to an initial state in which all unit cells are prepared in the doublon superposition state. Even though in this case the correlations start to spread further for $\theta=\pi/4$, they still remain more confined than for $\theta=0$ with the same initial state, and also even more than for $\theta=0$ with the initial state of Fig. \ref{ABcaging_MB} (a) and Fig. \ref{ABcaging_MB} (b).  This analysis indicates that the novel Aharonov-Bohm caging features of doublons in this model can be used to create (approximate) many-body cages, with potential uses and applications in atomtronic devices and circuits~\cite{amico2021roadmap}. 

\section{Conclusion}
\label{conclusion}
We have examined the properties of two- and three-boson bound states in the Creutz ladder. In the flat-band single-particle limit, the two-boson bound states admit a completely analytical description which we have used to identify and analyze the appearance of unusual two-body edge states formed by the bonding of a single-particle topological edge state and a localized flat-band state. We have also derived an effective model for doublons in the strongly interacting limit, showing behaviour similar to that of single particles. This effective model unveils the presence of non-topological doublon edge states which hybridize with the topological ones and tend to displace them one cell away from the boundaries of the ladder. 

When we futher extend this to three-body bound states in the regime of strong interactions, we again show that such trions replicate the properties of single particles and have non-topological edge states. However, due to their larger energy difference with the bulk these states have a lower hybridization with the topological edge states, which are therefore more clearly displaced to the unit cells next to the edges of the ladder. Finally, we examined Aharonov-Bohm caging of bound states. In the situation in which the single-particle band structure is flat, we have shown that a certain type of two-body bound states exhibit perfect Aharonov-Bohm caging for arbitrary values of the interaction strength. In the strongly interacting limit it is also possible to observe approximate Aharonov-Bohm of doublons by tunning the hopping phase. In this scenario, we have shown that a many-body state formed by multiple doublons localized in different parts of the ladder exhibits a clear suppression of the spreading of two-body correlations.  

We note that this approach can be generalized to any number of particles, offering a route to explore many-body topological phases in flat-band systems. 
\begin{acknowledgments}
This work was supported in part by the EPSRC Programme Grant DesOEQ (EP/P009565/1).
\end{acknowledgments}
\appendix 
\section{Schrieffer-Wolff trnasformation}
\label{AppendixSW}
In this appendix we provide a detailed derivation of the Schrieffer-Wolff transformation used in the main text. Let us consider an optical lattice filled with $N$ interacting bosons. We describe this system with a standard Bose-Hubbard Hamiltonian with hoppings $J_{ij}$ and a uniform onsite interaction strength $U$
\begin{equation}
\hat{H}=\frac{U}{2}\sum_i \hat{n}_i(\hat{n}_i-1) +\sum_{\langle i,j\rangle}J_{ij} \hat{a}_i^{\dagger}\hat{a}_j\equiv \hat{H}_U+\hat{H}_J.
\end{equation}
In the strongly interacting limit $U\gg J_{ij}$,  this model supports bound states of up to $N$ bosons that are well-separated in energy from the rest.  We are interested in deriving an effective model for these high-energy states. By treating the tunneling as a perturbation, it is possible to find such an effective model \cite{macdonald1988t} using a unitary transformation known as Schrieffer-Wolff (SW) transformation \cite{bravyi2011schrieffer}, which diagonalizes the Hamiltonian perturbatively in blocks of states with different onsite occupations. To proceed with the SW transformation, we decompose the kinetic term of the Hamiltonian $\hat{H}_J$ into a sum of terms that change the number of onsite bosonic pairs by a quantity $m$
\begin{equation}
\hat{H}_J=\sum_{m=-(N-1)}^{N-1} \hat{H}_J^m.
\end{equation}
Since each of the $\hat{H}_J^m$ terms creates (for $m>0$) or annihilates (for $m<0$) $m$ pairs of bosons, and the energy associated with each pair is $U$, these terms must obey the following commutation relations with the interacting part of the Hamiltonian $\hat{H}_U$
\begin{equation}
\left[\hat{H}_U,\hat{H}_J^m\right]=mU\hat{H}_J^m.
\label{CommRels}
\end{equation} 
The transformation of the Bose-Hubbard Hamiltonian $\hat{H}$ under any unitary $e^{i\hat{S}}$ (with $\hat{S}^{\dagger}=\hat{S}$) can be computed as a perturbative series of nested commutators
\begin{equation}
\tilde{H}=e^{i\hat{S}}\hat{H}e^{-i\hat{S}}=\hat{H}+\frac{[i\hat{S},\hat{H}]}{1!}+\frac{[i\hat{S},[i\hat{S},\hat{H}]]}{2!}+...
\label{transHam}
\end{equation}
From eqs. \eqref{CommRels} and \eqref{transHam}, it can be readily seen that the 0th order terms that couple states with different numbers of onsite occupation (i.e. ,$\hat{H}_J^m$ with $m\neq 0$) can be removed from the transformed Hamiltonian $\tilde{H}$ by choosing
\begin{equation}
i\hat{S}=\sum_{m=-(N-1),m\neq 0}^{m=N-1} \frac{\hat{H}_J^m}{mU}. 
\label{transformation}
\end{equation}
After applying this transformation, $\tilde{H}$ still contains terms of $\mathcal{O}(\frac{J_{ij}^2}{U})$ that couple subspaces with different number of onsite occupancies. It is possible to iteratively construct $k$th order transformations to remove terms of  $\mathcal{O}(\frac{J_{ij}^{k+1}}{U^k})$ \cite{macdonald1988t}, but for our purposes it is sufficient to use the transformation given by eq.  \eqref{transformation}.

In order understand more clearly how this general procedure allows us to derive effective models for bound states, let us consider explicitly the cases of bound states of $N=2$ bosons (doublons) and $N=3$ bosons (trions).\\
\subsection{Effective Hamiltonian for doublons}
In order to derive an effective Hamiltonian for bosonic doublons using the SW transformation,  we restrict the maximum occupancy of any site to $n_{i}^{\text{max}}=2$.  By doing so, we can decompose the identity operator as $\mathds{1}=\delta_{n_i,0}+\delta_{n_i,1}+\delta_{n_i,2}$. Multiplying $\hat{H}_J$ by $\mathds{1}$ from the left and the right, we can rearrange the kinetic term in the following way
\begin{align}
&\hat{H}_J=\hat{H}_J^{-1}+\hat{H}_J^{0}+\hat{H}_J^{+1}, \\
&\hat{H}_J^{-1}=\sum_{\langle i,j\rangle}J_{ij} \left( \delta_{n_i,1} \hat{a}_i^{\dagger}\hat{a}_j \delta_{n_j,2}\right) \nonumber, \\
&\hat{H}_J^{0}=\sum_{\langle i,j\rangle}J_{ij} \left[\left( \delta_{n_i,1} \hat{a}_i^{\dagger}\hat{a}_j \delta_{n_j,1}\right)+\left( \delta_{n_i,2} \hat{a}_i^{\dagger}\hat{a}_j \delta_{n_j,2}\right)\right] \nonumber, \\
&\hat{H}_J^{+1}=\sum_{\langle i,j \rangle}J_{ij} \left( \delta_{n_i,2} \hat{a}_i^{\dagger}\hat{a}_j \delta_{n_j,1}\right) \nonumber.
\end{align}
For instance, $\hat{H}_J^{-1}$ annihilates a bosonic pair by moving a particle from a site $i$ with $n_i=2$ ($n_i=1$ at the end of the process) to an adjacent site $j$ with $n_j=0$ ($n_j=1$ at the end of the process).  It can be readily checked that these terms obey the commutation relations \eqref{CommRels} for $m=0, \pm 1$. We now apply the transformation $i\hat{S}=\frac{1}{U}(\hat{H}_J^{+1}-\hat{H}_J^{-1})$ to the total Hamiltonian $\hat{H}=\hat{H}_U+\hat{H}_J$ to write the transformed Hamiltonian as
\begin{align}
\tilde{H}&=\hat{H}_J^0+\hat{H}_U+\frac{[\hat{H}_J^{+1},\hat{H}_J^{-1}]+[\hat{H}_J^{0},\hat{H}_J^{-1}]+[\hat{H}_J^{+1},\hat{H}_J^{0}]}{U}\nonumber\\
&+\mathcal{O}(U^{-2}).
\label{transDoublons1}
\end{align} 
To proceed further, we consider that the total number of particles in the lattice is $N=2$. By retaining in eq. \eqref{transDoublons1} only the terms that act non-trivially in the doubly-occupied subspace, we arrive at the following effective Hamiltonian for doublons
\begin{equation}
\hat{H}_\text{eff}^{N=2}=\hat{H}_U+\frac{1}{U}\hat{H}_J^{+1}\hat{H}_J^{-1}+\mathcal{O}(U^{-2}).
\label{EffHamDoublons}
\end{equation}
Furthermore, in $\hat{H}_J^{+1}\hat{H}_J^{-1}/U$ we only need to take into account the hopping processes where a doublon splits and then recombines again in the same site or an adjacent site. In the former case we will obtain an effective onsite energy shift, while the second case corresponds to an effective second-order hopping term that gives rise to doublon motion.
\subsection{Effective Hamiltonian for trions}
Following a procedure analogous to the case of doublons, in order to derive an effective Hamiltonian for trions we restrict the local occupancy to $n_{i}^{\text{max}}=3$ and express the identity operator as ${\mathds{1}=\delta_{n_i,0}+\delta_{n_i,1}+\delta_{n_i,2}+\delta_{n_i,3}}$. This allows us to decompose the kinetic term of the Hamiltonian into the following parts
\begin{align}
\hat{H}_J&=\hat{H}_J^{-2}+\hat{H}_J^{-1}+\hat{H}_J^{0}+\hat{H}_J^{+1}+\hat{H}_J^{+2}, \\
\hat{H}_J^{-2}&=\sum_{\langle i,j \rangle}J_{ij} \left( \delta_{n_i,1} \hat{a}_i^{\dagger}\hat{a}_j \delta_{n_j,3}\right) \nonumber, \\
\hat{H}_J^{-1}&=\sum_{\langle i,j \rangle}J_{ij} \left[\left( \delta_{n_i,1} \hat{a}_i^{\dagger}\hat{a}_j \delta_{n_j,2}\right)+\left( \delta_{n_i,2} \hat{a}_i^{\dagger}\hat{a}_j \delta_{n_j,3}\right)\right] \nonumber, \\
\hat{H}_J^{0}&=\sum_{\langle i,j \rangle}J_{ij} \left[\left( \delta_{n_i,1} \hat{a}_i^{\dagger}\hat{a}_j \delta_{n_j,1}\right)+\left( \delta_{n_i,2} \hat{a}_i^{\dagger}\hat{a}_j \delta_{n_j,2}\right)\right.\nonumber\\
&\left.+\left( \delta_{n_i,3} \hat{a}_i^{\dagger}\hat{a}_j \delta_{n_j,3}\right)\right] \nonumber, \\
\hat{H}_J^{+1}&=\sum_{\langle i,j \rangle}J_{ij} \left[\left( \delta_{n_i,2} \hat{a}_i^{\dagger}\hat{a}_j \delta_{n_j,1}\right)+\left( \delta_{n_i,3} \hat{a}_i^{\dagger}\hat{a}_j \delta_{n_j,2}\right)\right] \nonumber, \\
\hat{H}_J^{+2}&=\sum_{\langle i,j \rangle}J_{ij} \left( \delta_{n_i,3} \hat{a}_i^{\dagger}\hat{a}_j \delta_{n_j,1}\right) \nonumber.
\end{align}
For instance, $\hat{H}_J^{-2}$ annihilates two bosonic pairs by moving a particle from a site $i$ with $n_i=3$ ($n_i=2$ at the end of the process) to an adjacent site $j$ with $n_j=0$ ($n_j=1$ at the end of the process).  Again, these terms obey the commutation relations \eqref{CommRels} with $m=0,\pm 1, \pm 2$.  The SW transformation is now achieved with ${i\hat{S}=\frac{1}{U}\left(\frac{1}{2}\hat{H}_J^{+2}+\hat{H}_J^{+1}-\hat{H}_J^{-1}-\frac{1}{2}\hat{H}_J^{-2}\right)}$.  The general expression of the transformed Hamiltonian $\hat{H}$ becomes quite lengthy, but setting the total number of particles to $N=3$ and retaining only the terms that act non-trivially on the trion subspace, we arrive at the following minimal effective Hamiltonian allowing for trion motion
\begin{equation}
\hat{H}_\text{eff}^{N=3}=\hat{H}_U+\frac{1}{2U}\hat{H}_J^{+2}\hat{H}_J^{-2}+\frac{1}{4U^2}\hat{H}_J^{+2}\hat{H}_J^{0}\hat{H}_J^{-2}+\mathcal{O}(U^{-3}).
\end{equation}
In the second-order term $\hat{H}_J^{+2}\hat{H}_J^{-2}/2U$ we only need to take into account hopping processes were one particle forming the trions moves to an adjacent site and then recombines to the same site, giving rise exclusively to an energy shift. On the other hand, the third-order term $\hat{H}_J^{+2}\hat{H}_J^{0}\hat{H}_J^{-2}/4U^2$ may give rise both to processes in which a particle forming the trion hops three times and then recombines to the same site (energy shift), or in which the three particles forming the trion hop to an adjacent site (effective trion hopping).
\bibliography{main.bbl}

\end{document}